\documentclass[aps,pre,twocolumn,showpacs,superscriptaddress,amsmath,amssymb]{revtex4-1}
\usepackage{graphicx}
\usepackage{epsf}
\usepackage{bm}

\usepackage{txfonts}
\usepackage{color}

\begin{document}
\title{
Lane Formation Dynamics of Oppositely Self-Driven Binary Particles:\\
 Effects of Density and Finite System Size
}

\author{Kosuke Ikeda}
\affiliation{
Graduate School of Science and Technology, Niigata University, Niigata 950-2181, Japan
}

\author{Kang Kim}
\email{kk@cheng.es.osaka-u.ac.jp}
\affiliation{
Division of Chemical Engineering, Graduate School of Engineering
Science, \\
Osaka University, Toyonaka, Osaka 560-8531, Japan
}

\begin{abstract}
We examined the lane formation dynamics of oppositely self-driven binary
 particles by molecular dynamics simulations of a two-dimensional system.
Our study comprehensively revealed the effects of the density and system
 size on the lane formation.
The phase diagram distinguishing the no-lane and lane states
 was systematically
 determined for various combinations of the anisotropic friction
 coefficient and the desired velocity.
A peculiar clustered structure was observed when the lane was destroyed
 by considerably increasing the desired velocity.
A strong system size effect was demonstrated by the
 relationship between the temporal and spatial scales of the lane structure.
This system size effect can be attributed to an analogy with the driven
 lattice gas.
The transport efficiency was characterized from the scaling
 relation in terms of the degree of lane formation and the interface
 thickness between different lanes.
\end{abstract}

\maketitle

\section{Introduction}

A recent challenge in research on active matter systems has been
understanding 
the basic principles of various self-organizations observed in
 the collections of self-propelled or self-driven
 elements~\cite{Vicsek:2012gp, Marchetti:2013bp, Menzel:2014je,
 Bechinger:2016cf}.
Representative examples include schools of fish, flocks
of birds, swarms of animals, and crowds of humans.
In these systems, all elements exhibit macroscopic and synchronized
 collective behavior in
 the same direction despite microscopic repulsive interactions.
The self-driven force can lead to the spontaneous symmetry
 breaking of the locomotion, which is significantly different from
 passive Brownian motion.

Lane formation is a remarkable example of self-organization in
many autonomous agent systems, and is associated with pedestrian
counterflows~\cite{Helbing:1999cl, Helbing:2001fl, Castellano:2009ce}.
Specifically, 
when two groups of agents are driven and move in opposite directions, 
the agents of the same species tend to move collectively and eventually form
a lane in the driven direction.
Such segregation of the different lanes is closely related
to how efficiently the pedestrians can flow toward the desired
direction in various crowded
environments, such as evacuation in the event of a
disaster~\cite{Chowdhury:2005bn, Karamouzas:2014cz}.

The social force model was proposed to
obtain the underlying mechanism of a pedestrian flow from the
microscopic level~\cite{Helbing:1995bi, Helbing:2000dv}.
In this model, each pedestrian is modeled as a particle with a finite
excluded volume that interacts via repulsion to avoid collisions.
Additionally, the self-driven force acts on each particle in order to
adjust the instantaneous velocity to the preassigned desired direction and magnitude.
A large number of simulation studies have been carried out to
optimize realistic pedestrian
behavior~\cite{Helbing:2000ef, Yu:2005wt, Yu:2007em, Moussaid:2010gg, Zanlungo:2014gb}.

A similar lane-formation phenomenon has also been demonstrated in
simulations of binary oppositely charged
particles~\cite{Dzubiella:2002jh, Chakrabarti:2004kc,
Rex:2007cg, Rex:2008du, Liu:2008bg, Wysocki:2011jf, Jiang:2011fb,
Kohl:2012cy, Glanz:2012io, Lichtner:2014fy, Heinze:2015wr, Kogler:2015vw, Foulaadvand:2016iz,
Klymko:2016ce, Wachtler:2016fw} as well as
experiments~\cite{Leunissen:2005kc, Sutterlin:2009da, Vissers:2011gw, Vissers:2011je}.
The phase diagram of the lane formation has been
examined for various combinations of the external force and
the density under the periodic boundary
condition~\cite{Dzubiella:2002jh, Chakrabarti:2004kc, Rex:2007cg,
Rex:2008du, Liu:2008bg, Wachtler:2016fw} or with
confined geometry~\cite{Heinze:2015wr, Foulaadvand:2016iz}.
The finite-size effect has also been
investigated~\cite{Glanz:2012io, Heinze:2015wr, Foulaadvand:2016iz}.
In particular, the critical force that generates lane formation
exhibits a logarithmic increase with the system size, which implies that global
phase segregation cannot occur in a finite amount of time~\cite{Glanz:2012io}.
Recently, the observed system size dependence has been interpreted
through an analogy with the driven lattice gas model~\cite{Klymko:2016ce}.
That is, the lane formation in the driven system is basically governed by the phase separation.
However, global lane formation extending to the system size from the
disordered and random configuration would occur with infinite time.

With regard to the oppositely self-driven particles,
recent advances have been seen in simulations for the 
phase diagram of lane formation
with square system geometry under the periodic boundary condition~\cite{Ikeda:2012gd}.
In particular, the anisotropic friction perpendicular to the driven
direction has been introduced, which mimics the anisotropic response of
pedestrians after collisions.
However, the information regarding the density and finite-size effects
is still limited, in contrast to the data obtained in the above 
simulation studies on driven systems under an external field.

In this study, we examined the lane formation dynamics of oppositely
self-driven binary particles by simulating Langevin molecular dynamics
of a two-dimensional system.
In particular, we systematically examined the effects of the density and
finite size on lane formation.
Various order parameters regarding the lane structure were
quantified: the degree of lane formation, velocity fluctuation, transport
efficiency, characteristic time scale of lane formation, lane number,
lane width, and interface thickness between different lanes.
Such comprehensive quantitative analysis allowed us to clarify both the
similarity and differences between self-driven and 
external-field-driven lane formations.

This paper is organized as follows.
In Sect. 2, we introduce the simulation model and the numerical details.
In Sect. 3, we define the various observables to characterize the lane formation.
In Sect. 4, we present the numerical calculations.
Finally, in Sect. 5 we summarize the results.

\section{Simulation model and details}

We simulated the Brownian molecular dynamics
for two-dimensional equimolar binary mixtures of $N$ particles in a
regular square.
In the mode, 
each particle is assigned the variable $s=+$ or $-$.
Half of the particles have $s=+$, and the other half have $s=-$.
This variable $s$ characterizes the driven direction (along the $x$-axis) when a
self-driven force is applied to the particle, which is explained below.
The number density is $\rho= N/L^2$, where $L$ is the linear dimension
of the system.
The periodic boundary condition is used in both the $x$- and $y$-directions.
The equation of motion for the position $\bm{r}_i(t)$ and velocity
$\bm{v}_i(t)=d\bm{r}_i(t)/dt=(v_{ix}, v_{iy})$ of the $i$-th particle is given by
\begin{equation}
m\frac{d\bm{v}_i}{dt} = -\nabla_i \sum_{j\ne i} \phi(r_{ij})-\Gamma
 \bm{v}_i + \bm{R}_i(t) + \bm{F}^{\mathrm{(d)}}_i.
\end{equation}
Here, $m$ denotes the particle mass.
$\phi(r_{ij})$ is the interaction potential, which
depends only on the pair separation $r_{ij}=|\bm{r}_i-\bm{r}_j|$ between
particles $i$ and $j$.
The purely repulsive soft-core potential
$\phi(r)=\epsilon(\sigma/r)^{12}$ is used
with the cutoff length $r_c=2.5\sigma$.
Here, $\epsilon$ and $\sigma$
represent the magnitude of the potential energy and particle
diameter, respectively.
In addition, $\bm{R}_i(t)$ stands for the zero mean random force acting on the
$i$-th particle.
The variance of $\bm{R}_i(t)$ is given by $\langle
R_{i\alpha}(t)R_{j\beta}(t')\rangle= 2mk_BT \Gamma
\delta_{ij}\delta_{\alpha\beta}\delta(t-t')$, which is characterized by
the thermal energy 
$k_BT$ and the friction coefficient $\Gamma$ that should be associated with the
collisions with the solvent.
$\alpha$ and $\beta$ represent the two Cartesian components.

The self-driven force acting on particle $i$ is given by
\begin{equation}
\bm{F}^{\mathrm{(d)}}_i = -\gamma_x(v_{ix} - s_i V_0) \bm{e}_x - \gamma_y
 v_{iy} \bm{e}_y.
\end{equation}
Here, the first term denotes Helbing's proposed social force model related to pedestrian
phenomena~\cite{Helbing:2001fl}.
The $x$ component of the $i$-th particle
velocity tends to attain the desired velocity $V_0$ in the driven $x$-direction,
which is determined by the variable $s_i$.
The corresponding P\'{e}clet number $Pe$ is then described by $Pe =
\gamma_x V_0\sigma/k_BT$.
This model includes the relaxation time due to the friction $\gamma_x$
in the $x$-direction.
For the $y$-direction, on the other hand, the $y$-component of the velocity
decays to zero because of the friction $\gamma_y$ perpendicular to
the driven direction.
Here, the friction coefficients $\gamma_x$ and 
$\gamma_y$ are chosen independently to imitate the anisotropic
effect of the momentum exchange with the substrate~\cite{Ikeda:2012gd}.
The drift velocity of the single particle in the stationary state is
described by $\langle v^{s}_\mathrm{st}\rangle = s\gamma_x V_0/(\gamma_x+\Gamma)$.
If the random force $\bm{R}_i(t)$ and associated friction force $-\Gamma
\bm{v}_i$ are not taken into account, our model equation reduces to that used in the
previous study~\cite{Ikeda:2012gd}.
In addition, if the frictions $\gamma_x$ and $\gamma_y$ become zero,
the considered equation of motion reduces to the usual Langevin
equation of the many-particle system.

Hereafter, the units of length, temperature, and time
are represented as $\sigma$, $\epsilon/k_B$, and $\sqrt{m\sigma^2/\epsilon}$, respectively.
In all simulations, $m=1$, $T = 1$, and $\gamma_x=1$ were fixed values, and 
a time step of $\Delta t=0.001$ was used.
The dependence of 
the number density $\rho$ on the lane formation for various $V_0$ and
$\gamma_y$ was first intensively investigated at the fixed particle number $N=108$.
The examined densities were $\rho=0.2$, $0.3$, $0.4$, $0.5$, $0.6$,
$0.7$, $0.8$, $0.9$, and $1.0$.
Next, we focused on the finite-size effect on lane formation by
changing the particle number as $N=108$, $200$, $400$, $1000$, $4000$,
and $10000$ at the fixed number density $\rho=0.8$.
After equilibration with the thermal energy $k_BT=1$ using the
Langevin equation, the self-driven force $\bm{F}^{\mathrm{(d)}}$ was
applied to the particles.
In this study,
10 independent $t=10^3
(N=108)$ and $t=2\times 10^4$ ($N=200$, $400$, $1000$, $4000$, and $10000$)
trajectories were generated, from which various
quantities were calculated.

\begin{figure*}[t]
\centering
\includegraphics[width=.9\textwidth]{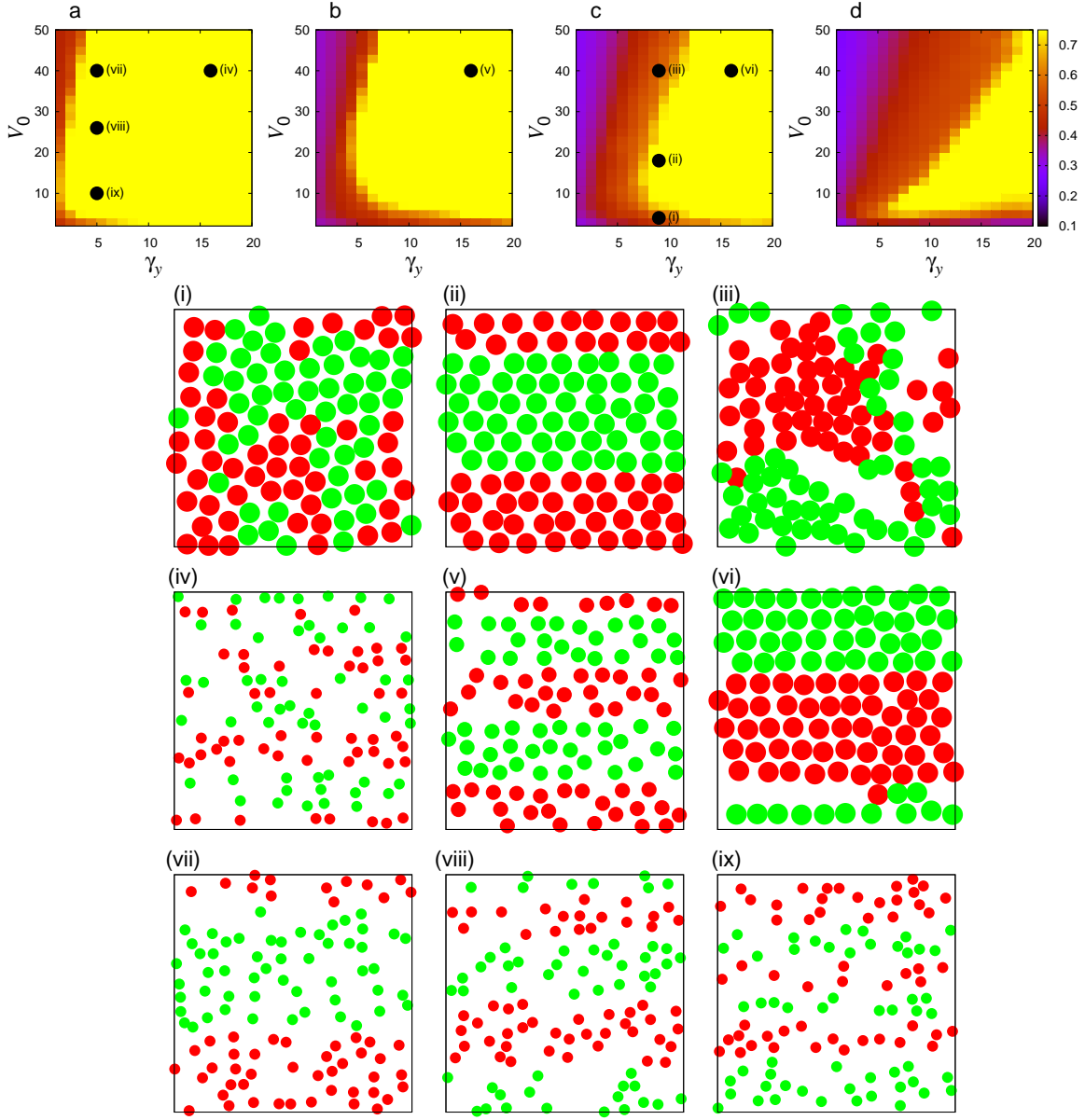}
\caption{(Color online)
Phase diagrams representing
the lane formation order parameter $\Psi$ for various combinations of
the friction along the $y$-direction $\gamma_y$ and the desired velocity
 $V_0$ at the densities $\rho=0.2$ (a), 0.5 (b), 0.8 (c), and 1.0 (d).
Typical snapshots of the states (i)-(ix) in the phase diagrams are also illustrated.
In these snapshots, note that the linear dimension of the system is
 normalized so that
 the particle size changes with the
 density.
}
\label{fig_op_lane}
\end{figure*}
\begin{figure*}[t]
\centering
\includegraphics[width=.9\textwidth]{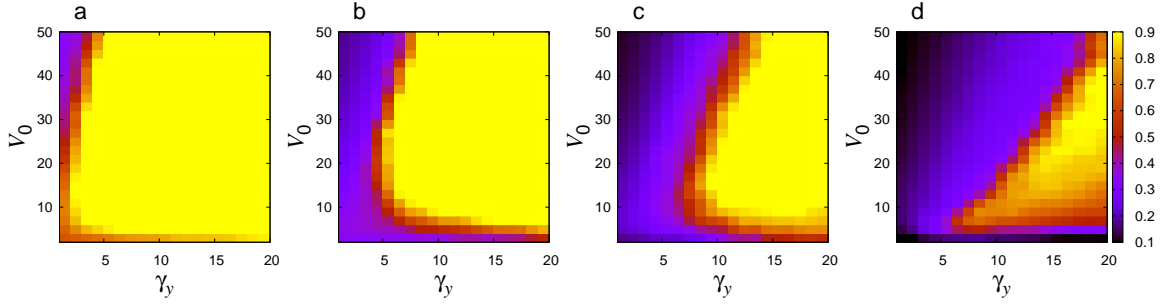}
\caption{(Color online)
Phase diagrams representing
the transport efficiency $\eta$ for various combinations of
the friction along the $y$-direction $\gamma_y$ and the desired velocity
 $V_0$ at the densities $\rho=0.2$ (a), 0.5 (b), 0.8 (c), and 1.0 (d).
}
\label{fig_efficiency}
\end{figure*}

\begin{figure*}[t]
\centering
\includegraphics[width=.9\textwidth]{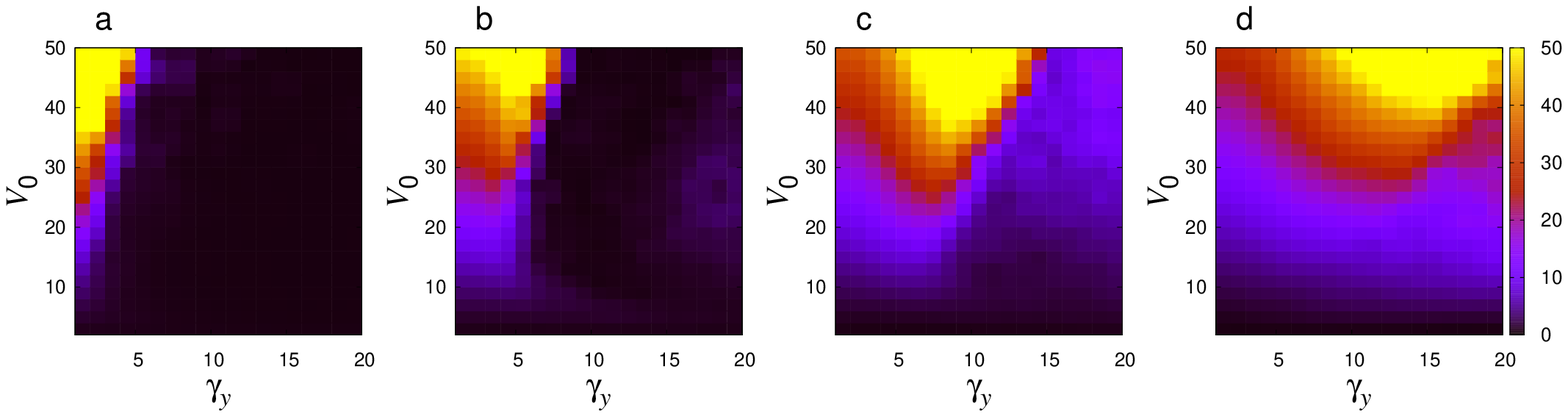}
\caption{(Color online)
Phase diagrams representing
the velocity fluctuation in the $x$-direction $\Delta$ for various combinations of
the friction along the $y$-direction $\gamma_y$ and the desired velocity
 $V_0$ at the densities $\rho=0.2$ (a), 0.5 (b), 0.8 (c), and 1.0 (d).
}
\label{fig_velocity_variance}
\end{figure*}

\section{Order parameters}
\label{order_parameter}

To characterize the degree of lane formation, some order parameters were
proposed for evaluating the number of particles of the same species along the driven direction.
As introduced in Ref.~\onlinecite{Ikeda:2012gd},
the simulation system can be virtually divided into identical rectangular strips
along the driven $x$-direction.
The number of strips is given by $n_\mathrm{div}=[L/\sigma]+1$, where $[\cdots]$
denotes the Gauss symbol.
Thus, the width of each strip along the $y$-direction is described by $\ell_\mathrm{div}=L
/n_\mathrm{div}$, which is closer to the particle diameter $\sigma$.
The order parameter for the lane formation in
the $k$-th rectangular strip is defined as
\begin{equation}
\psi_k = \frac{n_k^{+}-n_k^{-}}{n_k^{+}+n_k^{-}},
\end{equation}
where $n_k^{\pm}$ represents the number of $s=\pm$ species particles
in the $k$-th rectangular strip.
This signed variable $\psi_k$ is determined by the difference in numbers
between the positively and negatively driven particles $n_k^{+}-n_k^{-}$,
which is normalized by the
total particle number $n_k^{+}+n_k^{-}$ in the $k$-th strip.
Thus, the range of $\psi_k$ is from $-1$ to $1$, and 
$\psi_k$ becomes positive if 
the $s=+$ particles become dominant in the strip and negative if the 
$s=-$ particles become dominant.
The order parameter $\Psi$ for the total system is then defined
by
\begin{equation}
\Psi = \left\langle\frac{1}{n_\mathrm{div}}
		     \sum_{k=1}^{n_\mathrm{div}} |\psi_k|\right\rangle,
\end{equation}
where $\langle\cdots \rangle$ is the ensemble and time average.
The order parameter $\Psi$ begins with around zero at the initial state and
finally reaches 1 when the system exhibits the lane
formation in the steady state.
The former corresponds to the randomly mixed state, whereas the latter
corresponds to the lane formation structure by the particles of the same species
along the driven direction.

The transport property is depicted by the driven efficiency in
the $x$-direction, \textit{i.e.}, the average
drift velocity normalized by the single particle $\langle v^s_x\rangle$,
\begin{equation}
\eta = \frac{1}{2}\left(\frac{\langle\bar{v}^{+}_x\rangle}{\langle
		   v^{+}_\mathrm{st}\rangle} +
		   \frac{\langle\bar{v}^{-}_x\rangle}{\langle
		   v^{-}_\mathrm{st}\rangle}\right), 
\end{equation}
where $\bar{v}^{\pm}_x=(2/N)\sum_{i=1}^{N/2} v^{\pm}_{ix}$ is the average magnitude of the
particle velocity in each driven direction.
The velocity fluctuation in the $x$-direction is calculated from
\begin{equation}
\Delta = \frac{1}{2} \left(\langle\Delta^+\rangle + \langle\Delta^-\right\rangle),
\end{equation}
where $\Delta^{\pm} = (2/N)\sum_{i=1}^{N/2} (v^{\pm}_{ix} - \langle
\bar{v}^{\pm}\rangle)^2$ is the velocity variance in each driven direction.

The number of lanes $N_\mathrm{L}$ was also evaluated.
For this, 
the order parameter of each strip $\psi_k$ was first 
discretized with respect to the threshold values $\pm 0.5$ to
$\psi^\mathrm{d}_k= 1$ $(\psi_k > 0.5)$, $0$ $(|\psi_k| \le 0.5)$, and $-1$
$(\psi_k <-0.5)$.
Then, 
the nearest $l$-th strip for which the order parameter is $\psi^{\mathrm{d}}_l\ne 0$ from 
the $k$-th lane can be determined as follows:
\begin{equation}
l(k) = \min_{1<l-k<n_\mathrm{div}}[l + (k+n_\mathrm{div}-1)
 \delta_{0, \psi^{\mathrm{d}}_l}].
\end{equation}
Here, note that the maximum and minimum of the index $l$ are given by
$k+n_\mathrm{div}-1$ and $k+1$, respectively.
In addition, the periodic boundary condition
$\psi^\mathrm{d}_{j+n_\mathrm{div}} = \psi^\mathrm{d}_j$ was utilized.
Because the number of lanes $N_\mathrm{L}$ is equal to that of
the interface between the oppositely driven lanes, the following
equation can be obtained:
\begin{equation}
N_\mathrm{L} = \left\langle\sum_{k=1}^{n_\mathrm{div}} \delta_{-1, \psi^{\mathrm{d}}_k\psi^{\mathrm{d}}_{l(k)}}\right\rangle.
\end{equation}
The average thickness of the interface $\xi$ was estimated by
\begin{equation}
\xi = \left\langle\frac{\ell_\mathrm{div}}{N_\mathrm{L}}\sum_{k=1}^{n_\mathrm{div}}
(l(k)-k-1)\times \delta_{-1, \psi^{\mathrm{d}}_k\psi^{\mathrm{d}}_{l(k)}}\right\rangle,
\end{equation}
for $N_\mathrm{L}\ne 0$.
Correspondingly, the average lane width along the $y$-direction
$W_\mathrm{L}$ can be evaluated from the following relation:
\begin{equation}
\frac{L}{N_\mathrm{L}} = W_\mathrm{L} + \xi.
\label{eq_NL}
\end{equation}
Thus, the average length scale per lane
$L/N_\mathrm{L}$ can be evaluated from the combination of the lane width
$W_\mathrm{L}$ and the interface thickness $\xi$.

\begin{figure}[t]
\centering
\includegraphics[width=.35\textwidth]{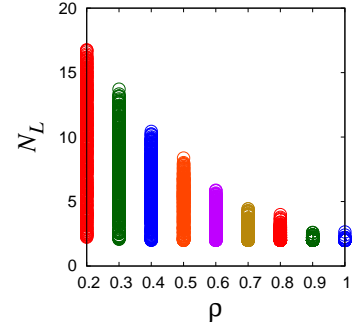}
\caption{(Color online)
Lane number $N_\mathrm{L}$ of the laned states ($\Psi\ge 0.8$)
 as a function of the density $\rho$.
}
\label{fig_lane_number_108}
\end{figure}

\begin{figure*}[t]
\centering
\includegraphics[width=.6\textwidth]{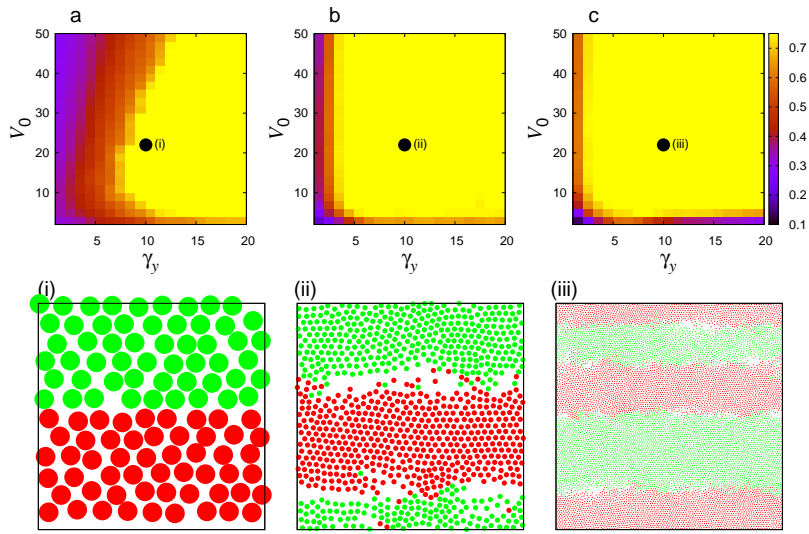}
\caption{(Color online)
Phase diagrams representing
the lane formation order parameter $\Psi$ for various combinations of
the friction along the $y$-direction $\gamma_y$ and the desired velocity
 $V_0$ at the density $\rho=0.8$ for the system sizes $N=108$ (a), 1000 (b), and
 10000 (c).
Typical snapshots of the state $(\gamma_y, V_0)=(10, 22)$
are illustrated as (i)-(iii).
In these snapshots, note that the linear dimension of the system is
 normalized so that the
 particle size changes with the
 density.
}
\label{fig_op_lane_fss}
\end{figure*}

\begin{figure}[t]
\centering
\includegraphics[width=.35\textwidth]{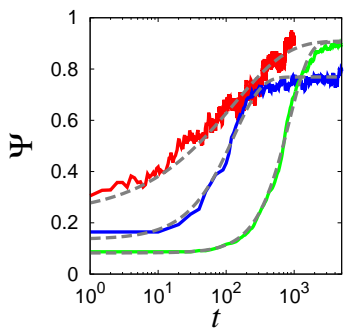}
\caption{(Color online)
Time development of the lane formation order parameter $\Psi$ with
 the particle number $N=108$ (red), $1000$ (blue), and $10000$ (green).
The state was set to $(\gamma_y, V_0)=(10, 22)$, which is the same as the
 snapshot in Fig.~\ref{fig_op_lane_fss}.
The dashed lines represent the fitting results with Eq.~(\ref{eq_psi}).
}
\label{fig_time_dependence}
\end{figure}

\section{Results and discussion}
\label{results}

\subsection{Density dependence on lane formation at $N=108$}
\label{results_N108}

First, we show the results of various observables in the system with $N=108$.
In particular, the density dependence is systematically discussed in this subsection.
Figure~\ref{fig_op_lane} plots the lane formation order
parameters $\Psi$ for various combinations of
$\gamma_y$ and $V_0$.
Snapshots of the states (i)-(ix) in the phase diagrams are also
illustrated.
We confirm that
the transition lines in phase diagrams eventually converged
at final stages in our simulations.
Thus, the utilized simulation time is sufficient to achieve the steady
states.
The characteristic time of lane formation will be discussed later.
These figures indicate that both large $V_0$ and $\gamma_y$ led
to lane formation.
With increasing $V_0$, the collision between oppositely driven particles was enhanced.
In addition, increasing $\gamma_y$ reduced the mobility in the $y$-direction
perpendicular to the driven direction.
These two effects resulted in a bundle of particles in the same
direction and generated the interface between oppositely driven
particles, as seen in the snapshots of Fig.~\ref{fig_op_lane}.
This was the main mechanism for the lane formation, which is in accord with
the previous study using Newtonian dynamics~\cite{Ikeda:2012gd}.
Figure~\ref{fig_efficiency} plots a contour map of the
transport efficiency $\eta$.
This indicates that the overall behaviors of $\Psi$
and $\eta$ were well correlated.
Namely, higher lane formation provided more efficient transportation.

Both Figs.~\ref{fig_op_lane} and \ref{fig_efficiency} demonstrate that
the region of the lane formation states became smaller as the density
increased.
This can be understood as follows.
The mean free path of the
particle decreased with increasing density.
Accordingly, a position exchange after a collision between oppositely driven particles
is unlikely to occur in such dense states.
In addition, the transition line between nonlane and lane states showed 
nonmonotonic and reentrant features, particularly at higher densities.
That is, the lane state reentered the no-lane state 
when $V_0$ was greatly increased at a fixed $\gamma_y$ [see Figs.~\ref{fig_op_lane}(i)-(iii)].
To visualize the details of the reentrance,
the corresponding 
velocity fluctuation in the driven $x$-direction $\Delta$ was analyzed,
as illustrated in Fig.~\ref{fig_velocity_variance}.
The no-lane states with larger $V_0$ exhibited markedly larger
velocity fluctuations at all densities.
Figure~\ref{fig_op_lane} shows a
typical snapshot at $\rho=0.8$ as state (iii).
The interface between opposite
lanes became unstable, and the same species tended to aggregate to form
a cluster.
Finally, the lane structure became a
clustered structure, where collisions of
oppositely driven clusters frequently occurred.

Note that the observed reentrance is not equivalent to
similar phenomena observed in the lane formation of externally driven
particles~\cite{Chakrabarti:2004kc}.
In the latter, the reentrant behavior between the no-lane and lane structures
demonstrated dependence on the density at some fixed external force.
Specifically, lane formation was unlikely because the thermal fluctuation
became dominant at low densities even with
a large external force.
In contrast, as seen in Figs.~\ref{fig_op_lane} and
\ref{fig_efficiency}, the transition line of the present self-driven
system monotonically changed as a function of density.
This was due to the anisotropic friction $\gamma_y$, which reduced the
fluctuation of the mobility in the $y$-direction even at low densities.

\begin{figure*}[t]
\centering
\includegraphics[width=.95\textwidth]{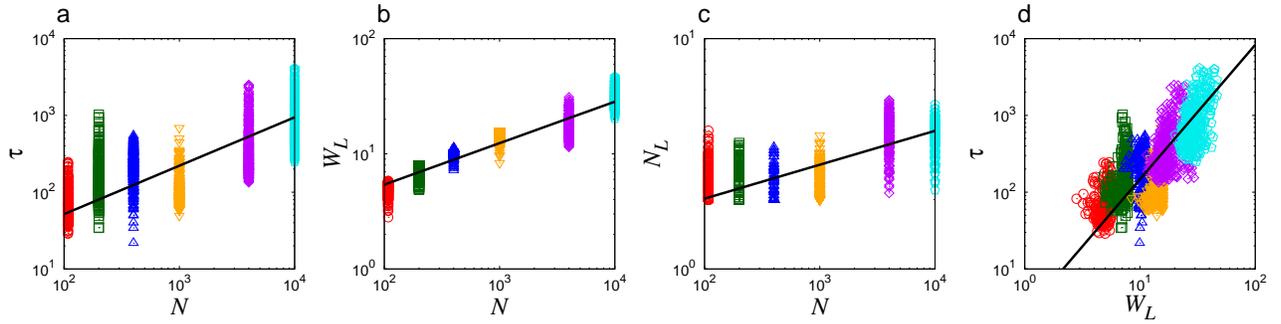}
\caption{(Color online)
Particle number dependence of the lane formation coarsening time
 $\tau$ (a), lane width $W_\mathrm{L}$ (b), and lane number
 $N_\mathrm{L}$ (c).
The states were chosen for $\Psi\ge 0.8$.
The straight lines represents the power laws $N^{0.63}$ (a), $N^{0.36}$ (b),
 and $N^{0.15}$ (c), 
which were determined by fitting for
the plotted data of each figure.
(d) Relationship between $\tau$ and $W_\mathrm{L}$.
The scaling $\tau\sim {W_\mathrm{L}}^{1.75}$ is represented
by a straight line.
}
\label{fig_NL_fss}
\end{figure*}

\begin{figure*}[t]
\centering
\includegraphics[width=.7\textwidth]{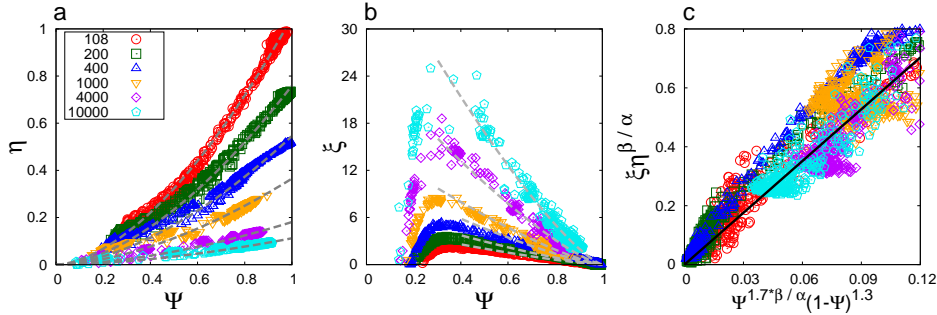}
\caption{
(a) Transport efficiency $\eta$ as a function of the lane formation
 order parameter $\Psi$ for various particle numbers $N$.
The dashed curves represent the fitting results using the function
 $\eta = AN^{-\alpha} \Psi^{1.7}$ with the constant $A=10.2$ and the
 exponent $\alpha=0.49$.
(b) Interface thickness $\xi$ as a function of the lane formation
 order parameter $\Psi$.
The dashed curves represent the fitting results using the function
 $\xi = BN^{\beta}(1- \Psi)^{1.3}$ with the constant $B=0.73$ and the
 exponent $\beta=0.44$.
Fitting was performed for $\Psi \ge 0.4$.
(c) Scaling relationship between $\Psi^{1.7\beta/\alpha}(1-\Psi)^{1.3}$ and
 $\xi \eta^{\beta/\alpha}$.
The states were chosen for $\Psi\ge 0.4$.
The straight line represents the proportional relationship with the slope
 of $A^{\beta/\alpha} B$.
}
\label{fig_fss_scaling}
\end{figure*}

The snapshots of states (iv)-(vi) in Fig.~\ref{fig_op_lane}
show that 
the number of lanes $N_\mathrm{L}$ apparently depended on the
density $\rho$.
Figure~\ref{fig_lane_number_108} presents a 
scatter plot of $N_\mathrm{L}$ with various $V_0$ and $\gamma_y$ as
a function of the density $\rho$.
This indicates that $N_\mathrm{L}$ gradually decreased with increasing
$\rho$.
As mentioned above, lane formation is caused by the collision
between oppositely driven particles.
At high densities, where the mean free path becomes smaller, the collisions
of oppositely driven particles are correlated to one another; then,
the aftereffects
are plausibly widespread to some extent to make a wider
bundle of same-species particles.
This leads to a smaller number of lanes at higher densities.
The range of $N_\mathrm{L}$ was observed to become progressively wider
when the density decreased.
The snapshots of states (vii)-(ix) ($\gamma_y=16$, $V_0=40$) in
Fig.~\ref{fig_op_lane} show the strong dependence of $N_\mathrm{L}$ on $V_0$ with
a fixed $\gamma_y=5$ at the density $\rho=0.2$.
In fact, 
the fluctuation of the mobility in the $y$-direction became smaller with
a larger $\gamma_y$.
This resulted in a larger $N_\mathrm{L}$ because the
collisions were less correlated than those at higher densities.
In contrast, both larger $V_0$ and smaller $\gamma_y$ reduced
$N_\mathrm{L}$ even at low densities, for which
the mechanism was similar to that at higher densities.
This effect led to the broadly scattered distribution at lower densities,
as seen in Fig.~\ref{fig_lane_number_108}.
An analogous density dependence of the lane number was previously
reported for the external-field-driven systems~\cite{Liu:2008bg}.
In the study, however, the particle number was controlled to change the density in the
fixed simulation area.
This means that the system size effect has not been adequately investigated in
a systematic manner.
To overcome this issue,
we explored the details of the system size effect on lane formation, as described
in the next subsection.

\subsection{Finite-size effect on lane formation}
\label{results_fss}

In this subsection, we discuss the finite-size effect on the lane formation of
the oppositely self-driven particles.
Figure~\ref{fig_op_lane_fss} shows the system size dependence of the 
lane formation order parameter $\Psi$.
Apparently, the extent of the capability to form the laned structure increased
with the system size $N$.
In particular, the clustered structure that was exhibited for larger $V_0$ in the
$N=108$ system eventually disappeared for larger $N$ systems at any
state in the phase diagram.
The snapshots in Fig.~\ref{fig_op_lane_fss} show
that the geometric structures of the lane such as the lane number $N_\mathrm{L}$,
the lane width $W_\mathrm{L}$, and the interface thickness $\xi$, were affected by the
system size even when the parameters $\gamma_y$ and $V_0$ had the same values.

For more details, 
we examined 
the coarsening process of the lane formation with
the time evolution of $\Psi(t)$.
The averaged results of the state at $\gamma_y=10$ and $V_0=22$
are shown in Fig.~\ref{fig_time_dependence} for $N=108$, $1000$, and $10000$.
The characteristic time scale associated with the lane formation
from the randomly mixed configuration became slower when the system size
increased.
This lane formation time scale is quantified by the
following function:
\begin{equation}
\Psi(t) = \Psi_\mathrm{in} + (\Psi_\mathrm{st}-\Psi_\mathrm{in}) \left[1-e^{-(t/\tau)^{c}} \right],
\label{eq_psi}
\end{equation}
where $\Psi_\mathrm{in}$ and $\Psi_\mathrm{st}$ are given by the lane formation order parameter
at the initial and steady states, respectively.
Thus, the fitting parameters are the exponent $c$ and the characteristic
time $\tau$.
The results are plotted in Fig.~\ref{fig_time_dependence}.
Figure~\ref{fig_NL_fss}(a) plots the quantified time scale $\tau$
as a function of $N$ for the state $\Psi\ge 0.8$.
The characteristic time of the coarsening $\tau$
scaled as $\tau\propto N^{0.63} (\propto L^{1.26})$ on average.
This implies that the macroscopic lane formation occurs at the
thermodynamic limit.
A comparable system size dependence has been reported for the lane formation
caused by the external field~\cite{Glanz:2012io}.
Furthermore, 
analogous scaling has been demonstrated in the coarsening
dynamics of the driven lattice-gas model~\cite{Levine:2001jq,
Hurtado:2003hi, Saracco:2003ke, Ohta:2012be}.
In addition, 
driven on- and off-lattice model
simulations have recently been compared~\cite{Klymko:2016ce}.

The system size dependences of the lane width $W_\mathrm{L}$ and the
lane number $N_\mathrm{L}$ are plotted
in Figs.~\ref{fig_NL_fss}(b) and (c), respectively.
If the system size effect is negligible, the lane width $W_\mathrm{L}$ should converge
to some finite value; then, $N_\mathrm{L}$ increases according to
$N_\mathrm{L}\propto L(\propto\sqrt{N})$.
However, as mentioned above, the lane formation can only exist with infinite time.
Correspondingly, the lane width 
approximately followed $W_\mathrm{L}\propto N^{0.36} (\propto L^{0.72})$ on
average, which is weaker than the proportional relation to the system linear dimension $L$.
Indeed,
Fig.~\ref{fig_NL_fss}(c) indicates that the lane number $N_\mathrm{L}$ slightly
increased with the particle number $N$.
In other words, the two-lane state with the width
$L/2$ could not develop in the finite time simulations.
The relationship between the length and time scales
$\tau\sim {W_\mathrm{L}}^{1.75}$ was eventually obtained, as shown in Fig.~\ref{fig_NL_fss}(d).
These numerical results can be understood as the dynamical effect of the
collisions of opposite particles.
The two-lane state is energetically the most stable because the interface effect of the
opposite driven particles is lowest.
However, when a lane
with the width $W_\mathrm{L}=L/2$ is formed from the disordered
initial configuration,
the number of collisions between different species increases with
the system size.
To reduce such energy loss due to collisions during the
coarsening process, the lane should be split to increase 
the lane width $W_\mathrm{L}\propto L^{0.72}$, as
shown in Fig.~\ref{fig_NL_fss}(b).

Finally, the system size effect of the transport efficiency $\eta$ was examined.
Figure~\ref{fig_fss_scaling}(a) plots the relationship between $\Psi$ and
$\eta$ for various particle numbers $N$.
As discussed in Sect.~\ref{results_N108}, the lane formation order
parameter $\Psi$ correlated with $\eta$ at $N=108$.
However, the overall efficiency $\eta$ decreased even at the larger $\Psi$
state for $N=10000$.
This dependence of $\eta$ on $\Psi$ can be described by the proportional
relation as $\eta = AN^{-\alpha}\Psi^{1.7}$ with the constant $A\simeq 10.2$ and
the exponent $\alpha\simeq 0.49$.
As noted above, particles of different species within the interface
collided frequently, which reduced the transport efficiency $\eta$.
Indeed, the number of the interface slightly increased with $N_\mathrm{L}$,
as shown in Fig.~\ref{fig_NL_fss}(c).
However, the observed reduction of $\eta$ with increasing $N$ was rather substantial.
To clarify this phenomenon, the interface thickness $\xi$ was plotted as a
function of the degree of the lane formation $\Psi$, 
as shown in Fig.~\ref{fig_fss_scaling}(b).
The interface thickness $\xi$ increased with a higher particle number $N$.
According to the definition given in Eq.~(\ref{eq_NL}), $\xi$ becomes 
zero at the nonlane state $\Psi=0$ with $N_\mathrm{L}=0$.
As shown in the figure, 
the peak was remarkably pronounced at the intermediate ordered state $\Psi \simeq 0.3$.
Furthermore, $\xi$ strongly decreased toward zero when $\Psi$ approached unity.
This reduction behavior can be described by
$\xi = BN^{\beta} (1-\Psi)^{1.3}$ with the constant $B\simeq 0.73$ and
the exponent $\beta\simeq 0.44$.
From the two equations, the particle number $N$ dependence can be
deleted; thus, the relationship among the three variables $\Psi$,
$\xi$, and $\eta$ is given by $\Psi^{1.7 \alpha/\beta}(1-\Psi)^{1.3} \sim \xi
\eta^{\alpha/\beta}$, as presented in Fig.~\ref{fig_fss_scaling}(c).

\section{Summary}

We numerically studied the lane formation dynamics of
oppositely self-driven particles based on the social force model.
In particular, we addressed the density and finite-size effects on
the lane formation process.

First, the density dependence of the lane formation order parameter
$\Psi$ was analyzed.
The transition lines that distinguish the no-lane and lane states can be
described by various combinations of the desired velocity $V_0$
and the anisotropic friction coefficient $\gamma_y$.
The density dependence of the transition line monotonically changes,
\textit{i.e.}, the area of the lane formation region decreases with
increasing density.
Such monotonic behavior is contrary to the observed lane
formation caused by an external field, where the reentrant effect
is exhibited in the density dependence of the critical external force
for lane formation~\cite{Chakrabarti:2004kc}.
This discrepancy is associated with the friction $-\gamma_y v_y$, which reduces the
fluctuation in the direction perpendicular to the driven direction.
Thus, the ability of the lane formation in the present model becomes
higher than that in Ref.~\onlinecite{Chakrabarti:2004kc}.
Alternatively, we found a different type of reentrant structure in the
dynamical phase diagram, particularly at the higher density state.
If the desired velocity $V_0$ becomes excessively large, the interface
of different lanes
becomes unstable and forms clustered structures of same-species particles.

Second, the finite-size effect on the lane formation was
systematically analyzed.
In particular, the system size dependence of the spatiotemporal scales
of the lane formation was characterized.
The coarsening process of the lane is analogous to the driven
lattice-gas model, where the macroscopic two-lane state is realized at
the thermodynamic limit.
By contrast, with finite simulation times,
the dynamic effects due to collisions of oppositely driven
particles, particularly within the interfaces, cause the lane to split.
The system size dependence of the transport efficiency was
quantified with two variables: the lane formation order parameter and the
interface thickness of different lanes.
The obtained scaling relation allows us to predict how efficiently 
particles flow toward a desired direction by knowing the degree of 
lanes and interface thickness for any system size.

\begin{acknowledgments}
The authors thank H. Hayakawa and H. Wada for helpful discussions.
K.I. is grateful to Y. \={O}no and N. Matubayasi for supporting
 his stay at Osaka University.
This work was partially supported by 
JSPS KAKENHI Grant Numbers JP26400428 on Scientific Research (c) 
and JP16H00829 on Innovative Area
 (2503) ʻʻStudying the Function of Soft Molecular Systems by the
 Concerted Use of Theory and Experimentʼʼ.
The computations were performed at Research Center of Computational
Science, Okazaki Research Facilities, National Institutes of Natural
 Sciences, Japan.
\end{acknowledgments}

\end{document}